\documentclass[final, 5p, times]{elsarticle}
\DeclareUnicodeCharacter{2113}{\ensuremath{\ell}}
\usepackage{subcaption}
\usepackage{hyperref}
\usepackage{xcolor}
\usepackage{amsmath}
\journal{arxiv.org}

\begin{document}
\begin{frontmatter}

\title{Reactor-based Search for Axion-Like Particles using CsI(Tl) Detector}


\author[inst1]{S. Sahoo}
\author[inst1]{S. Verma}
\author[inst1]{M. Mirzakhani}

\author[inst1]{N. Mishra}
\author[inst1]{A. Thompson}
\author[inst1]{S. Maludze}
\author[inst1]{R. Mahapatra}
\author[inst1]{M. Platt}

\affiliation[inst1]{organization={Department of Physics and Astronomy, Texas A\&M University },
            addressline={578 University Dr}, 
            city={College Station},
            postcode={77840}, 
            state={TX},
            country={US}}
            

\begin{abstract}
The absence of conclusive signals in weakly interacting massive particle (WIMP) searches has motivated increased interest in alternative dark matter candidates such as axions and axion-like particles (ALPs), which also provide a solution to the strong CP problem. In this work, we employ a $\sim100~\mathrm{kg}$ scale CsI(Tl)-based detector operated in proximity to a nuclear reactor to achieve a sub-100 DRU (differential rate unit, expressed in counts/keV/kg/day) background level in the MeV energy range through a combination of active veto and passive shielding techniques. Such a low-background environment enables sensitivity to ALPs with axion--photon coupling $g_{a\gamma\gamma} \gtrsim 10^{-6}$ and axion--electron coupling in the range $10^{-8} < g_{aee} < 10^{-4}$ for ALP masses between 1~keV and 10~MeV. These results demonstrate that the experiment has the potential to probe previously unexplored regions of parameter space, including the so-called cosmological triangle in the ALP--photon coupling for MeV-scale ALPs.
\end{abstract}
\end{frontmatter}


%
%
%
%
%

\section{Introduction}

Astrophysical observations including galaxy cluster dynamics, galactic rotation curves, and gravitational lensing, provide compelling evidence for the existence of non-luminous matter in the universe, commonly referred to as dark matter \cite{dent,Zwicky1,Zwicky2,Rubin1,alcock}. Among the proposed dark matter candidates, axions are particularly well motivated, as they not only constitute a viable dark matter candidate but also provide an elegant solution to the strong CP problem of the Standard Model.

Extensive experimental efforts have explored axion and axion-like particle (ALP) parameter space across a wide range of masses and couplings, employing different strategies depending on the source of ALPs. Helioscope experiments, such as CAST~\cite{zioutas,anastassopoulos,irastorza}, probe solar axions, while haloscope experiments including ADMX~\cite{asztalos,du}, HAYSTAC~\cite{Brubaker,droster}, Abracadabra~\cite{kahn,salemi}, and CASPEr~\cite{kimball} search for axions from the galactic dark matter halo and are primarily sensitive to low-mass ALPs ($m_a \lesssim 1~\mathrm{eV}$). Laboratory-based searches rely on the production of ALPs from intense photon sources. These include light-shining-through-walls experiments such as ALPS II~\cite{Spector}, as well as interferometric approaches like ADBC~\cite{liu} and DANCE~\cite{obata,melissinos,derocco}, which are also sensitive to light ALPs in the sub-eV mass range. In contrast, beam-dump and fixed-target experiments—including FASER~\cite{feng}, LDMX~\cite{berlin,aakesson}, NA62~\cite{volpe}, SeaQuest~\cite{berlin2018dark}, and SHiP~\cite{alekhin}—probe heavier ALPs, while hybrid proposals such as PASSAT~\cite{bonivento} combine multiple detection strategies. Direct dark matter detection experiments, including XENON~\cite{aprile,dent2020inverse}, SuperCDMS~\cite{aralis2020constraints}, and PandaX~\cite{fu2017limits}, also provide constraints on ALP couplings.

Despite this broad experimental program, a significant region of parameter space particularly in the MeV mass range remains relatively unexplored.
Reactor-based experiments provide a unique opportunity to probe this region due to the intense flux of $\gamma$ rays produced in nuclear fission processes. These photons can generate ALPs via Primakoff conversion, yielding a high-statistics source of MeV-scale ALPs in a controlled laboratory environment. When combined with low-background scintillation detectors, such as CsI(Tl), this approach enables sensitivity to ALP couplings that are complementary to existing constraints. In particular, the high photon flux and scalable detector mass allow reactor-based searches to explore previously inaccessible regions of ALP parameter space, including areas relevant to cosmological and astrophysical considerations.

This paper is organized as follows. Section 2 covers the theoretical framework of the experiment, detailing the production of ALPs at the reactor, their detection by our detector, and the overall sensitivity of the process. Section 3 outlines the experimental setup near the reactor, providing information about the reactor, detector, data acquisition process, sources of background, and different shielding techniques. In section 4, the results of the experiments are analyzed.Then in the last section we have added the exclusion plot for ALPs search in the desired region. It also go through the future improvements for the experiment and extending its reach to other unexplored region of the parameter space. 

\section{Theoretical Framework} \label{Theory}

This experiment is carried out at a TRIGA (Training, Research, Isotopes, General Atomics)-type nuclear reactor located at the Nuclear Science Center (NSC) at Texas A\&M University. The reactor produces an intense flux of photons through the fission of uranium in its core. These photons can subsequently interact with the reactor materials and generate axion-like particles (ALPs) via processes such as Primakoff conversion \cite{dent}, providing a high-statistics source of ALPs.

\begin{figure}[h!]
  \centering
    \includegraphics[width=\linewidth]{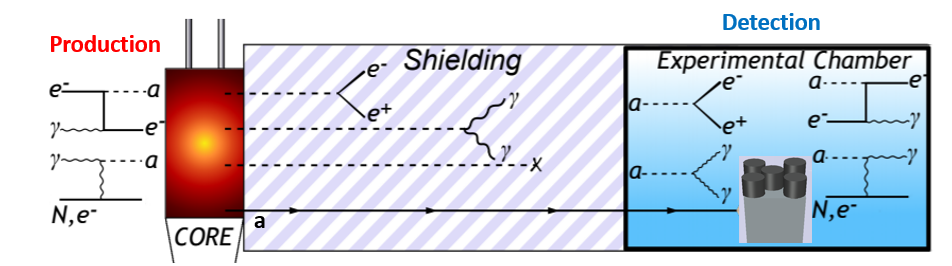}
   \caption{Schematic showing the experimental strategy for the production and detection of ALPs at the reactor. ALPs are generated at the reactor core via scattering with reactor material and detected using a CsI(Tl) detector assembly, as shown in the schematic. The ALP could decay and evade detection inside the shielding, depending on their lifetime (dashed lines) \cite{dent}. }
  \label{fig:expt_axion}
\end{figure}

Once produced, the ALPs can propagate through the shielding and reach the detector. If they interact within the CsI(Tl) scintillator, they can produce observable signatures such as photons or electrons through processes including inverse Primakoff scattering, Compton-like interactions, or decay into photon pairs or electron--positron pairs. These secondary particles are then detected in the CsI(Tl) crystals, enabling a search for ALP-induced signals \cite{Verma}. Figure~\ref{fig:expt_axion} shows a schematic illustration of ALP detection in the experiment.

\subsection{ALP Production Mechanism at the Reactor} \label{Alp_Production}

Fission, neutron capture in the fuel and surrounding materials, decay of fission products, inelastic scattering in the fuel, and the decay of activation products collectively produce an intense flux of photons in the reactor core \cite{roos}. These photons can subsequently interact with the reactor material, primarily $^{235}$U in our case, leading to the production of axion-like particles (ALPs) through several mechanisms: Primakoff conversion, Compton-like scattering, and nuclear de-excitation processes. For a detailed discussion of ALP production via nuclear de-excitation, we refer the reader to Refs.~\cite{sierra,waites}.

\begin{figure}[h!]
  \centering
  \includegraphics[width=\linewidth]{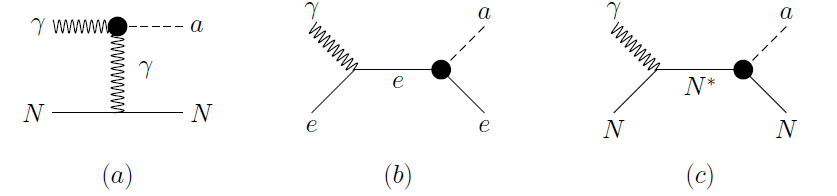}
  \caption{Mechanisms for producing ALPs at the reactor facility: (a) Primakoff process, (b) Compton-like process, and (c) nuclear de-excitation \cite{sierra}.}
  \label{fig:alp_production_feynmann}
\end{figure}

Figure~\ref{fig:alp_production_feynmann} illustrates these photon-induced ALP production mechanisms at the reactor: (a) Primakoff conversion ($\gamma + A \rightarrow a + A$, where $A$ denotes the atomic target), (b) Compton-like scattering ($\gamma + e^{-} \rightarrow a + e^{-}$), and (c) nuclear de-excitation ($\gamma + N \rightarrow N^{*} \rightarrow a + N$).

In this work, we consider a generic framework in which ALPs are produced via couplings to photons or electrons, and we evaluate the experimental sensitivity to ALP--photon and ALP--electron interactions within this setup \cite{Verma}.

\subsection{ALP Detection Mechanism at the Reactor}

Once produced, ALPs can be detected at the experimental site through several interaction channels. They can convert into photons via the inverse Primakoff process, $a + A \rightarrow \gamma + A$, where $A$ denotes the atomic target. ALPs can also interact with electrons through an inverse Compton-like process, $a + e^{-} \rightarrow \gamma + e^{-}$, leading to photon production. In addition, depending on their lifetime and mass, ALPs may decay within the detector volume into two photons or into an $e^{-}e^{+}$ pair. These detection channels are illustrated in Fig.~\ref{fig:expt_axion}.

The detector is enclosed within hermetic shielding to suppress external gamma-ray backgrounds. Furthermore, nuclear absorption processes, $a + N \rightarrow N^{*} \rightarrow \gamma + N$, can also provide a detection signature for ALPs at the detector site \cite{Verma}.

\subsection{Expected ALP Signal Analysis}

To a large extent, we adopt a minimal ALP framework in which the ALP--SM couplings governing production and detection arise from the same interaction vertices \cite{dent}. As illustrated in Fig.~\ref{fig:alp_production_feynmann}(a), the Primakoff process proceeds via the exchange of a virtual photon in the $t$-channel, with a rate determined by the ALP--photon coupling $g_{a\gamma\gamma}$ \cite{dent}. The differential cross section for ALP production via the Primakoff process in the reactor core is given by

\begin{equation} \label{eq:prim_prod}
    \frac{d \sigma^{p}_{P}}{d \cos\theta} = \frac{1}{4} g^{2}_{a\gamma\gamma} \alpha Z^{2} F^{2}(t)\,
    \frac{|\bar{p}_a|^{4} \sin^{2}\theta}{t^{2}} \, ,
\end{equation} where $t = (p_1 - k_1)^2 = m_a^2 + E_\gamma \left(E_a - |\bar{p}_a| \cos\theta \right)$ is the square of the four-momentum transfer. Here, $Z$ is the atomic number of the target nucleus, $\alpha$ is the fine-structure constant, $F(t)$ is the nuclear form factor, $|\bar{p}_a|$ denotes the magnitude of the outgoing ALP three-momentum, and $E_\gamma$ is the incident photon energy. The production cross section is enhanced by the coherent scattering factor $Z^2$, which significantly improves the sensitivity of the experiment to ALP production via the Primakoff channel compared to other interaction channels.

Similarly, the differential cross section for the inverse Primakoff process follows the same functional form as in Eq.~\ref{eq:prim_prod}, with the prefactor modified from $1/4$ to $1/2$. This difference arises from the change in the number of initial-state spin degrees of freedom, since the inverse process involves a spin-0 ALP instead of a spin-1 photon in the initial state.

ALPs can also decay into two photons within the detector volume. The corresponding decay width is given by

\begin{equation} \label{eq:photon_decay}
    \Gamma (a \rightarrow \gamma\gamma) = \frac{g_{a\gamma\gamma}^2 m_a^3}{64\pi} \, .
\end{equation} this decay width, when combined with the ALP kinetic energy, determines the decay length, which in turn governs the probability of the ALP decaying within the detector volume.

The model adopted in this work is intentionally minimal and focuses on ALP production via the Primakoff process and detection through inverse Primakoff conversion or decay into Standard Model particles. A more comprehensive treatment would include all production mechanisms discussed in Sec.~\ref{Alp_Production}, as well as additional detection channels involving more complex interactions with Standard Model (SM) particles. Such an extended analysis is beyond the scope of the present work and will be considered in future studies, reader might follow Ref.\cite{waites}, \cite{sierra} for the details.

Having established the differential cross sections for ALP production and detection processes, we now evaluate the expected event rates within this simplified framework. In this approach, ALPs are produced in the reactor core via the Primakoff process and are subsequently detected either through inverse Primakoff scattering or via their decay within the detector volume.

The total number of observed events is given by the sum of contributions from these detection channels \cite{sierra}:
\begin{equation} \label{ALP_count}
    N_{\text{Total}}^{P} = N_{P}^{P} + N_{D}^{P} \, ,
\end{equation}
where $N_{\text{Total}}^{P}$ denotes the total number of events for ALPs produced via the Primakoff process, $N_{P}^{P}$ corresponds to events detected through inverse Primakoff scattering, and $N_{D}^{P}$ represents events detected through ALP decay within the detector volume.

The detection probabilities for both channels are computed below, and these results are used to estimate the experimental sensitivity in the ALP parameter space.

Let's first consider the case of Primakoff ALPs production and inverse-Primakoff ALPs detection. To estimate the number of ALP signals, it is useful to factorize the process into three sequential components: (i) production of ALPs in the reactor core, (ii) propagation of ALPs from the core to the detector, and (iii) detection via inverse Primakoff scattering inside the detector. The equation is taken from Ref.\cite{Verma, sierra}. The resulting event rate though this process can be expressed as
\begin{equation}
    \text{Rate} = (\gamma\text{ flux}) \times (\text{ALP production}) \times  (\text{survival}) \times (\text{detection})
    \label{eq:rate}
\end{equation}

Hence, the total number of events is 

\begin{equation*}
    N = R \  \Delta t \\
\end{equation*}

\begin{equation} \label{eq:alp_detection_1}
N_{P}^P = 
\left(\int \frac{dE_\gamma}{(4\pi l_d^2)}  \,
\frac{dN_\gamma}{dE_\gamma} \,  \cdot
\frac{\sigma_P^p}{\sigma_{\text{SM}}}  \cdot
P_{\text{surv}} \,  \cdot m_{\text{det}} \, N_T \, \sigma_P^d\right) \, \Delta t \,
\end{equation}

In this expression, $m_{\text{det}}$ is the detector mass, $\Delta t$ is the data-taking time, and $N_T$ is the number of target atoms per unit mass of the detector material. The quantity $\sigma_P^d$ denotes the inverse Primakoff cross section governing ALP detection. The term $1/(4\pi l_d^2)$ accounts for the geometric dilution of the ALP flux as it propagates from the reactor core to the detector located at a distance $l_d$. The integral represents a convolution over the reactor photon spectrum, where $\frac{dN_\gamma}{dE_\gamma}$ is the differential photon flux in the core. The ratio $\sigma_P^p/\sigma_{\text{SM}}$ can be interpreted as the effective branching fraction for ALP production via the Primakoff process relative to all competing Standard Model photon interactions in the reactor medium. The factor $P_{\text{surv}}$ denotes the survival probability of ALPs propagating from the reactor to the detector without interacting, given by
\begin{equation} \label{eq:alp_surv_prob}
P_{\text{surv}} = \exp\left(-\frac{l_d}{v_a \tau_a}\right) \, ,  \tau_a = \frac{E_a}{m_a} \, \frac{1}{\Gamma(a \rightarrow \gamma\gamma)} \, 
\end{equation}
where $v_a$ is the ALP velocity and $\tau_a$ is the boosted ALP lifetime in the laboratory frame,

Similarly, for ALPs produced via the Primakoff process and detected through their decay into two photons, the expected event rate can be written (following Eq.~\ref{eq:rate}) as

\begin{equation} \label{eq:alp_detection_2}
    N_{D}^{P} = \left( \frac{A}{4\pi l_d^2} \int dE_\gamma \,
    \frac{dN_\gamma}{dE_\gamma} \,
    \frac{\sigma_P^p}{\sigma_{\text{SM}}} \,
    P_{\text{surv}} \,
    P_{\text{decay}} \right) \Delta t \, ,
\end{equation}

where $A$ is the transverse area of the detector. The factor $A/(4\pi l_d^2)$ accounts for the fraction of ALPs emitted from the reactor that geometrically intersect the detector. The integral represents the convolution over the reactor photon spectrum, with $\frac{dN_\gamma}{dE_\gamma}$ denoting the differential photon flux. The ratio $\sigma_P^p/\sigma_{\text{SM}}$ corresponds to the effective branching fraction for ALP production via the Primakoff process. The quantity $P_{\text{decay}}$ denotes the probability that an ALP decays within the detector volume and is given by

\begin{equation} \label{eq:alp_decay_prob}
    P_{\text{decay}} =
    \left[1 - \exp\left(-\frac{\Delta l}{v_a \tau_a}\right)\right] \, ,
\end{equation} where $\Delta l$ is the fiducial detector length. 

The sensitivity projections are computed using Eqs.~\ref{ALP_count}, \ref{eq:alp_detection_1}, and \ref{eq:alp_detection_2}, following the methodology described in Ref.~\cite{sierra}. A detailed feasibility study of the experimental approach is presented in Ref.~\cite{dent}. 

Motivated by these studies, a reactor-based axion search experiment was conducted at the Nuclear Science Center facility at Texas A\&M University during 2021--2022. The experiment was led by Shubham Verma. Further analysis of the collected data is currently ongoing toward the publication of the results.

\section{Experimental Setup}

This section describes the experimental configuration, including the reactor, detector system, background sources (both internal and external), shielding strategy, and the data acquisition procedure.

\subsection{Reactor}

\begin{figure}[h!]
    \includegraphics[scale=0.5]{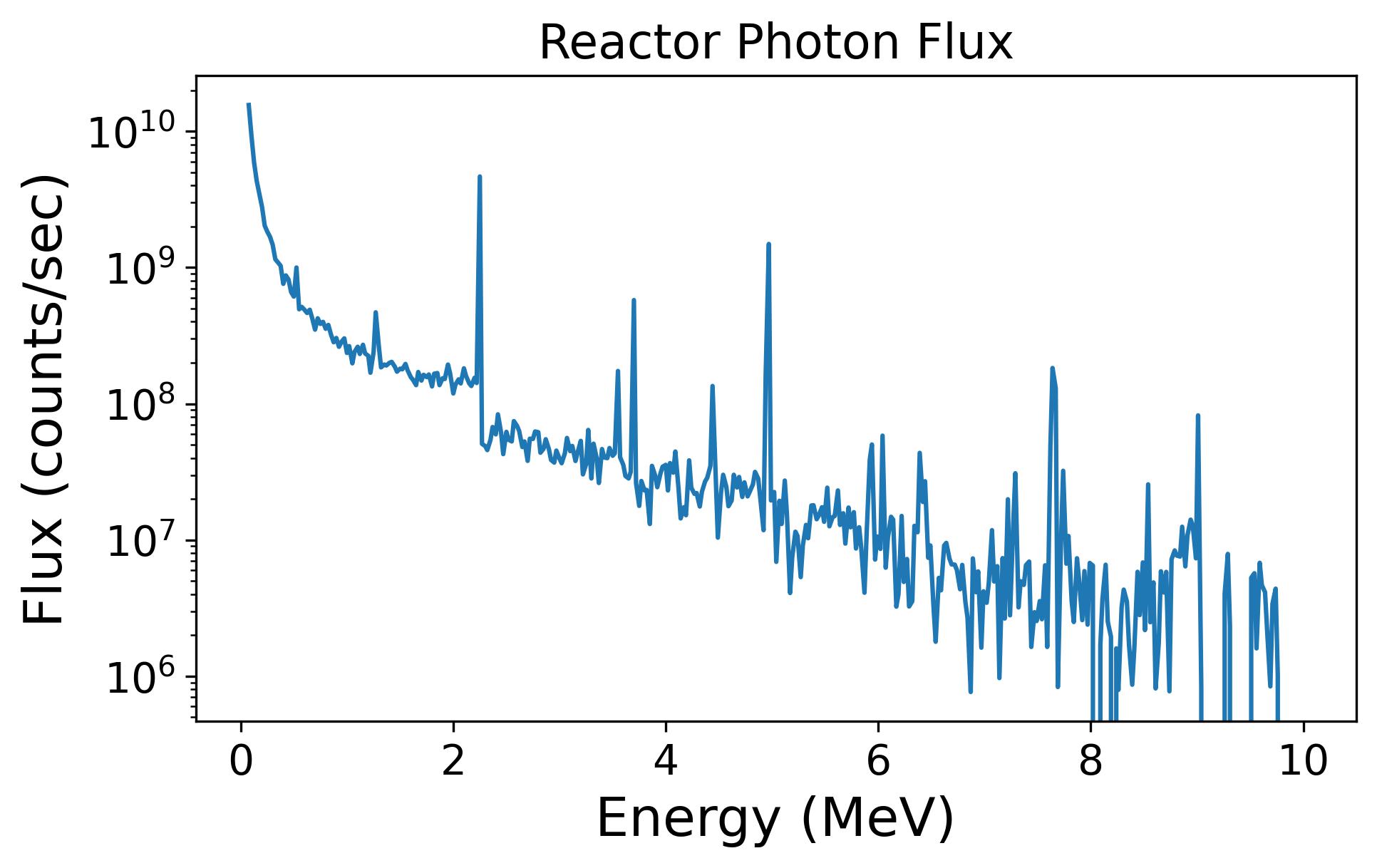}
   \caption{Photon flux coming from the reactor. }
  \label{fig:Flux_Reactor_Photon}
\end{figure}

As discussed in Section~\ref{Theory}, the experiment utilizes a 1~MW TRIGA research reactor. The reactor consists of a low-enriched uranium core operating at a thermal power of 1~MW. It produces an intense photon flux of approximately $9\times10^{11}~\text{photons}/\text{cm}^2/\text{s}$ through fission-induced processes \cite{minercollaboration2016background}. The simulated photon flux spectrum from the reactor is shown in Fig.~\ref{fig:Flux_Reactor_Photon} taken from Ref.\cite{minercollaboration2016background}. An important advantage of this facility is the movable core configuration, which allows adjustment of the detector-to-core distance. The detector can be positioned as close as 2~m from the reactor core, significantly enhancing the available photon flux at the experimental location.

\subsection{Detector}
\begin{figure}[h!]
\centering
\begin{subfigure}{0.24\textwidth}
  \centering
  \includegraphics[width=\linewidth]{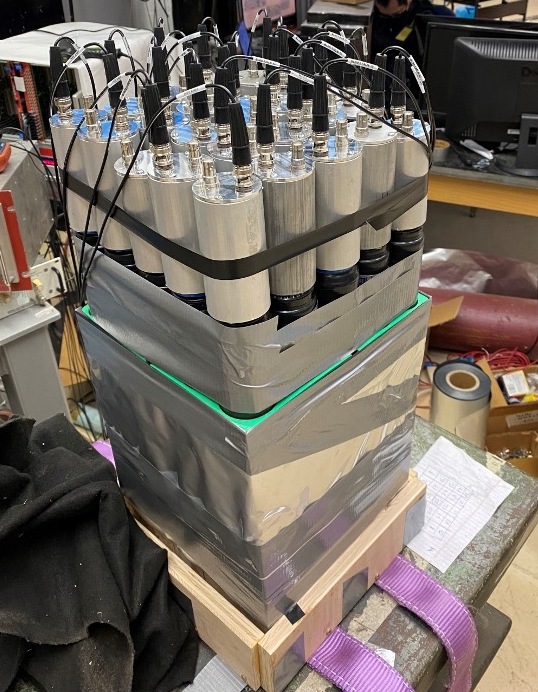}
  \caption{Crystal array and PMT coupling}
  \label{fig:sub-first}
\end{subfigure}
\hfill
\begin{subfigure}{0.22\textwidth}
  \centering
  \includegraphics[width=\linewidth]{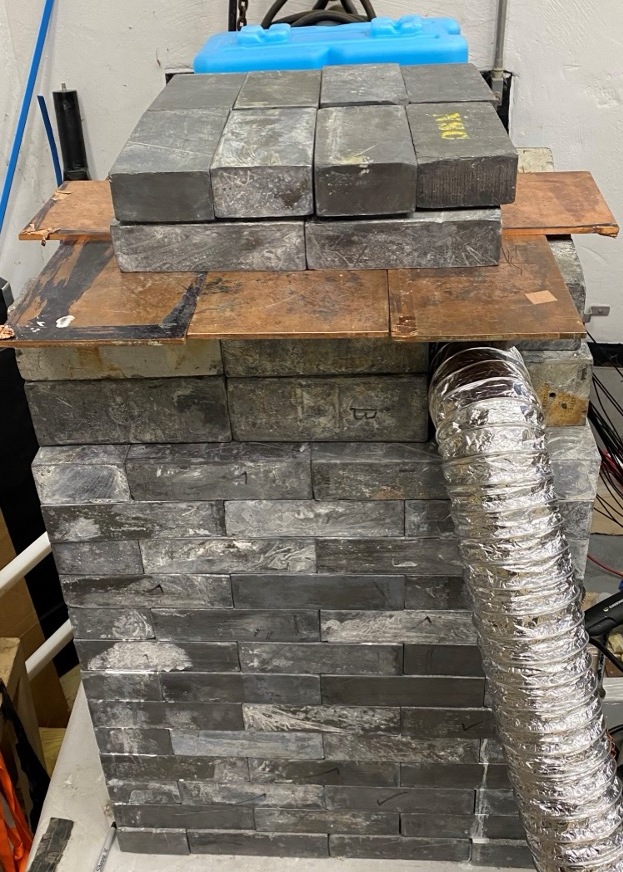}  
  \caption{Detector inside shielding}
  \label{fig:sub-second}
\end{subfigure}
\caption{CsI(Tl) detector assembly used in the experiment. (a) Packaging of the $5\times5$ array of crystals coupled to photomultiplier tubes (PMTs), including bases and high-voltage and signal connections. (b) Completed detector assembly enclosed within $4^{\prime\prime}$ thick lead shielding. The visible air tube provides cooling for the PMTs.}
\label{fig:Detector}
\end{figure}

The proposed experimental setup employs 25 CsI(Tl) scintillation crystals, each with a mass of 3.52~kg and dimensions of $2^{\prime\prime} \times 2^{\prime\prime} \times 12^{\prime\prime}$. The crystals are arranged in a compact $5 \times 5$ matrix configuration, as shown in Fig.~\ref{fig:Detector}. CsI(Tl) is selected as the scintillator material due to its high light yield ($\sim 5.4 \times 10^{4}$ photons per MeV of deposited $\gamma$-ray energy), relatively high density, and low hygroscopicity compared to other commonly used scintillators~\cite{pereira2018characteristics}. These properties make it well suited for detecting $\gamma$ rays produced in axion-like particle (ALP) searches via the Primakoff process, where the interaction rate benefits from the $Z^{2}$ enhancement discussed in Eq.~\ref{eq:prim_prod}~\cite{dent}. In addition, the low hygroscopicity of CsI(Tl) enables stable operation under standard laboratory conditions without the need for stringent environmental control.

Each CsI(Tl) crystal is wrapped with polytetrafluoroethylene (PTFE) tape to enhance the collection of scintillation light through improved internal reflection, with the emitted photons having a peak wavelength near 550~nm. A thin layer of aluminized Mylar is subsequently applied to provide optical isolation from external light~\cite{sahoo2026searchdarkmatterinvisible}. One end of each crystal is optically coupled to a conventional EMI photomultiplier tube (PMT) for efficient light collection~\cite{Verma}. A PMT is a highly sensitive vacuum detector that converts incident photons into an amplified electrical signal via the photoelectric effect followed by electron multiplication across multiple dynode stages (Ref.\cite{hamamatsu_pmt}). A potential divider base is attached to the PMT to supply stable bias voltages to the dynode chain. The combination of the CsI(Tl) crystal, PMT, and associated base forms a single detector module, and these modules are arranged in a $5 \times 5$ configuration to construct a $\sim 100~\mathrm{kg}$-scale detector for rare-event searches, as shown in Fig.~\ref{fig:Detector}.

\begin{figure}[h!]
\centering
\begin{subfigure}{0.45\textwidth}
  \centering
  \includegraphics[width=\linewidth]{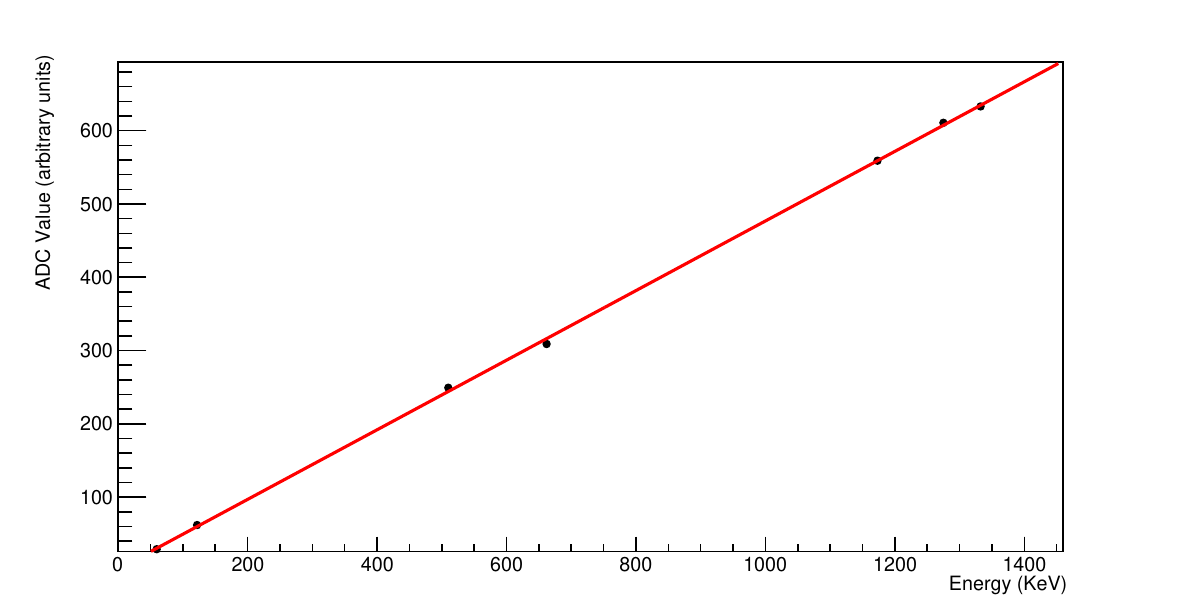}
  \caption{Energy calibration}
  \label{fig:CsI_calibration_plot}
\end{subfigure}
\hfill
\begin{subfigure}{0.45\textwidth}
  \centering
  \includegraphics[width=\linewidth]{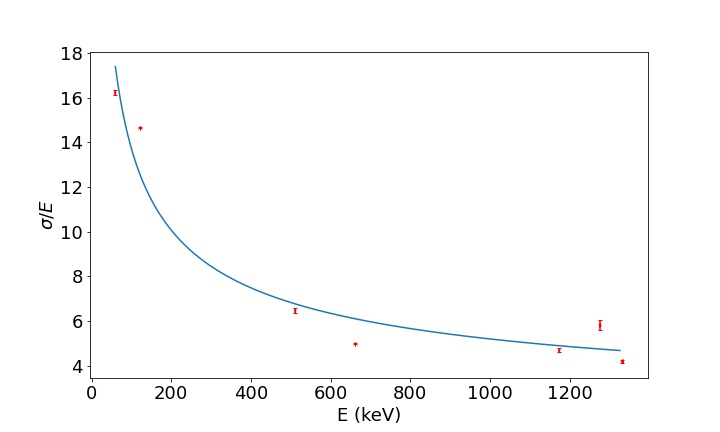}
  \caption{Energy resolution}
  \label{fig:CsI_energy_resolution}
\end{subfigure}
\caption{Performance characterization of CsI(Tl) crystals. (a) Calibration curve showing the linear relationship between ADC response and $\gamma$-ray energy. (b) Energy resolution ($\sigma/E$) as a function of energy, demonstrating improved resolution at higher energies.}
\label{fig:calibration_resolution}
\end{figure}

Various radioactive $\gamma$-ray sources, including $^{22}$Na, $^{137}$Cs, $^{60}$Co, $^{57}$Co, and $^{241}$Am, were used to characterize the response of the CsI(Tl) crystals employed in this experiment. These sources provide well-known $\gamma$-ray energies spanning a wide range, enabling precise calibration of the detector. Figure~\ref{fig:CsI_calibration_plot} shows the calibration curve obtained by correlating the measured ADC response with the known $\gamma$-ray energies. A clear linear relationship is observed over the energy range from $\sim$60~keV to 1.3~MeV, demonstrating the linearity of the CsI(Tl) detector response. This linear behavior is essential for accurate energy reconstruction and confirms the suitability of CsI(Tl) crystals for $\gamma$-ray detection in this energy range.

The lowest detectable energy threshold for the full-size CsI(Tl) crystal is found to be approximately 60~keV, as determined using the 60~keV $\gamma$-rays from a $^{241}$Am source (Ref.\cite{Verma}). This relatively low threshold is important for probing low-energy signals relevant to ALP searches. 

The energy resolution of the detector, defined as $\sigma/E$, is shown in Fig.~\ref{fig:CsI_energy_resolution}. As expected, the resolution improves with increasing energy, following the characteristic behavior governed by photon statistics and scintillation light yield. The measured resolution is consistent with the expected theoretical trend for CsI(Tl) scintillators \cite{Verma}.

\subsection{Backgrounds and Shielding} \label{subsec:backgrounds_shielding}

As the present work involves a rare-event search, a thorough understanding and mitigation of background radiation using the compact CsI(Tl) detector setup is essential for achieving the required sensitivity to axion-like particles. The dominant background contributions arise from multiple sources. External backgrounds include natural radioactivity in the surrounding environment, cosmic-ray interactions, and reactor-related radiation such as $\gamma$ rays, $\beta$ particles, neutrons, and neutron-induced radioactivity. Internal backgrounds originate from radioactive impurities within the detector materials, including the CsI(Tl) crystals and shielding components.

A detailed study of these background sources and their expected contributions is presented in Appendix~\ref{sec:Background_Appendices}. Effective suppression of these backgrounds is critical for improving the signal-to-noise ratio and enhancing the sensitivity to rare axion-induced events. To this end, both passive and active shielding techniques are employed.

\begin{figure}[h!]
\centering
	\includegraphics[scale=0.6]{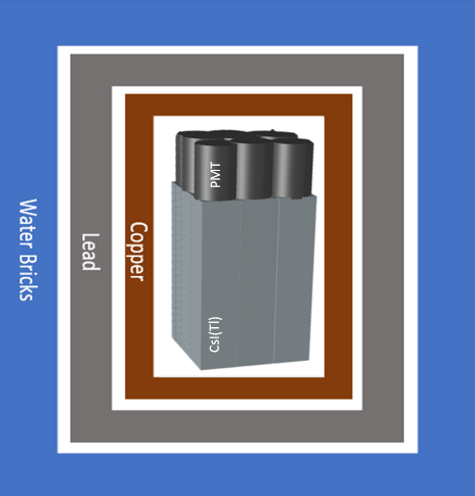}
	\caption{Schematic of the $3\times3$ prototype assembly with different layers of shielding. The detector is surrounded by passive shielding composed of $4^{\prime\prime}$ of lead, $1/4^{\prime\prime}$ of copper, and water bricks to reduce environmental background radiation~\cite{Verma}.}
	\label{fig:csi_shielding}
\end{figure}

Passive shielding was implemented using $1/4^{\prime\prime}$ thick copper, $4^{\prime\prime}$ thick lead, and water bricks, as illustrated in Fig.~\ref{fig:csi_shielding}. The shielding configuration was based on studies reported in Ref.~\cite{minercollaboration2016background}. Subsequent measurements with the detector indicated that the copper layers and water bricks had a negligible impact on the background level in the region of interest, spanning approximately 100~keV to a few MeV. Consequently, in later experimental runs, only the $4^{\prime\prime}$ thick lead shielding was retained. All lead bricks used for the shielding were individually tested using a Geiger counter to assess their intrinsic radioactivity. Only bricks exhibiting relatively low activity were selected for the final shielding configuration surrounding the detector assembly.

\begin{figure}[h!]
\centering
	\includegraphics[width=\linewidth]{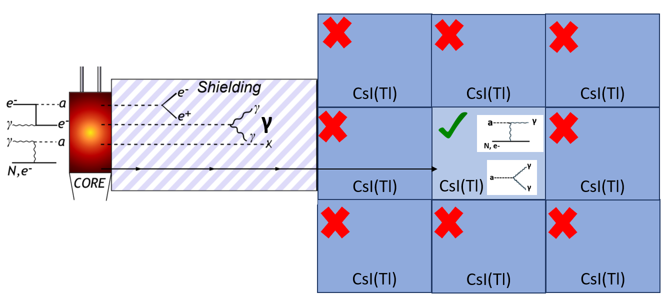}
	\caption{Illustration of the self-shielding or single-scatter veto technique using the outer layer of CsI(Tl) detectors for background reduction in the $3\times3$ prototype configuration~\cite{Verma}.}
	\label{active_veto}
\end{figure}

In addition to passive shielding, an active veto strategy was implemented using the outermost layer of CsI(Tl) detectors in the $5\times5$ array. This method is referred to as the anti-coincidence cut in the analysis presented throughout this work. The anti-coincidence or single-scatter cuts are applied during offline data analysis.

Figure~\ref{active_veto} illustrates the single-scatter technique for a $3\times3$ detector configuration. In this approach, the central detector is designated as the primary detector, while the eight surrounding detectors serve as active veto detectors. For each event recorded in the central detector, a coincidence window of $10~\mu$s is applied, chosen to accommodate the relatively long scintillation decay time of CsI(Tl), which is approximately $3~\mu$s~\cite{hawrami2021growthevaluationimprovedcsitl}. If a signal is observed in any of the surrounding veto detectors within this time window, the event is rejected as background, since genuine ALP-induced interactions are expected to deposit energy only in the primary detector. Conversely, events with no coincident signals in the veto detectors are accepted as candidates and included in the single-scatter energy spectrum.

The same procedure is extended to the $5\times5$ detector configuration, where the inner $3\times3$ array is treated as the primary detection region. In this case, the energies recorded in the nine central detectors are summed, effectively increasing the active detector mass, while the outer layer of detectors is used to apply the single-scatter veto condition.

\subsection{Data Acquisition}

High-voltage biasing for the photomultiplier tubes (PMTs) is provided by a CAEN SY5527LC mainframe equipped with two A7236 high-voltage modules, each capable of delivering up to 3.5~kV with a maximum current of 1.5~mA across 24 channels. The high-voltage system can be controlled remotely using the CAEN GECO2020 software.

Data acquisition is performed using a VME-based CAEN V1740D digitizer, a 64-channel, 12-bit analog-to-digital converter operating at 62.5~MHz. The digitizer supports data transfer rates of up to 80~MB/s via an optical link and is equipped with DPP-QDC (Digital Pulse Processing -- Charge-to-Digital Converter) firmware. In combination with the CAEN CoMPASS software, this setup enables real-time processing of input pulses to extract integrated charge (energy) and precise trigger time stamps for each event. This approach eliminates the need to store full raw waveforms, thereby significantly reducing data storage requirements. Additionally, the system supports remote data acquisition, allowing monitoring and operation of the experiment during reactor ON periods without requiring continuous on-site presence.

The experiment commenced in early 2021, during which detector testing and characterization were carried out. Data taking was completed by the end of 2022, yielding approximately 2 days of reactor ON data and 7 days of reactor OFF data. Since reactor operation was not continuous, the reactor ON dataset was accumulated over multiple shorter runs, which were subsequently combined to obtain the total exposure. Reactor OFF data were recorded a few hours after shutdown to minimize contributions from short-lived radioactive activation products. The reactor OFF spectrum is treated as the background and is used to estimate the statistical significance of any observed signal.

\section{Results and Analysis}

The experiment was developed in stages, beginning with a prototype setup consisting of 9 CsI(Tl) crystals arranged in a compact $3\times3$ geometry, and subsequently expanded to a $5\times5$ configuration with 25 crystals. In the initial $3\times3$ configuration, the central crystal is used to identify single-scatter events, while the eight surrounding crystals serve as an active veto. This defines a fiducial mass of 3.5~kg, corresponding to the mass of a single CsI(Tl) crystal. In the upgraded $5\times5$ configuration, the inner $3\times3$ array defines the fiducial volume, corresponding to a total fiducial mass of 31.5~kg, while the outer 16 crystals act as a veto layer.

\subsection{Prototype $3\times3$ CsI(Tl) Setup}

\begin{figure}[h!]
  \centering
    \includegraphics[scale=0.4]{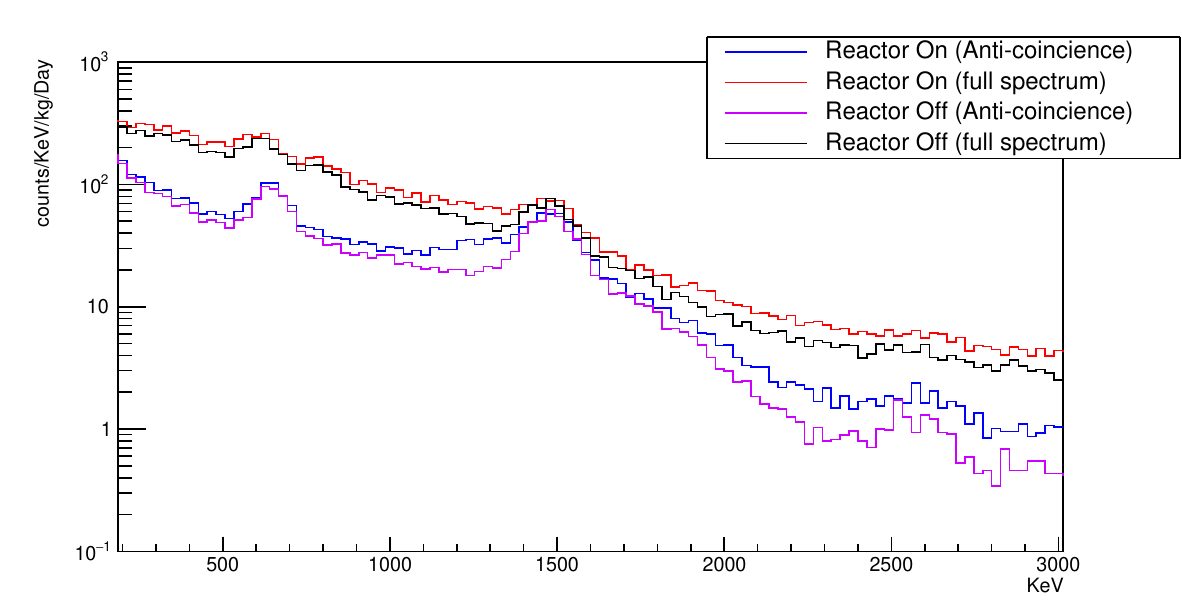}
   \caption{Comparison of reactor-on and reactor-off data for the central detector in the $3\times3$ configuration, with and without the anti-coincidence (single-scatter) cut~\cite{Verma}.}
  \label{fig:reactor_on_off_comparison_2}
\end{figure}

Data were collected using the $3\times3$ prototype setup with the full shielding configuration described in Sec.~\ref{subsec:backgrounds_shielding}. The detector was positioned at a distance of approximately 4~m from the reactor core to optimize the expected signal rate. The resulting energy spectra for reactor ON and reactor OFF conditions are shown in Fig.~\ref{fig:reactor_on_off_comparison_2}. Particular attention is given to the anti-coincidence spectrum, obtained after applying both active and passive background suppression techniques. The spectrum shows higher reactor ON compared to OFF signal as expected. 

To further reduce background contributions, an additional air-purge system was implemented around the $3\times3$ prototype assembly. The system consists of a $3/16^{\prime\prime}$ thick acrylic enclosure surrounding the detector, with openings for cabling and ventilation. Fresh air is introduced from the top via a compressor system at the NSC facility and circulated downward through the enclosure. The air-purge system is installed outside the lead shielding and is designed to suppress airborne radioactive contaminants, in particular $^{41}$Ar. This isotope is produced via neutron activation in the reactor environment and emits a characteristic $\gamma$ ray at 1.29~MeV. An excess in the background spectrum near 1.29~MeV is observed during reactor-on conditions.

\begin{figure}[h!]
  \centering
    \includegraphics[width=\linewidth]{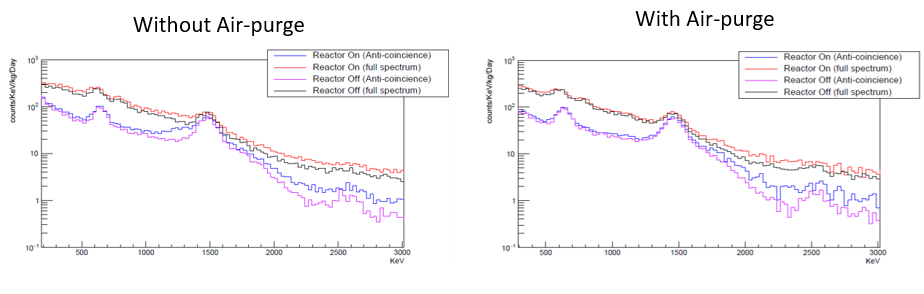}
   \caption{(Left) Reactor-on and reactor-off comparison for the $3\times3$ setup without air purge, showing an excess around 1.29~MeV due to $^{41}$Ar. (Right) Same comparison with the air-purge system applied, demonstrating a reduction of the 1.29~MeV background~\cite{Verma}.}
  \label{airpurge_2}
\end{figure}

The effectiveness of the air-purge system is demonstrated in Fig.~\ref{airpurge_2}, where a significant reduction in the 1.29~MeV $\gamma$-ray background is observed when the air purge is applied.

\begin{figure}[h!]
  \centering
    \includegraphics[width=\linewidth]{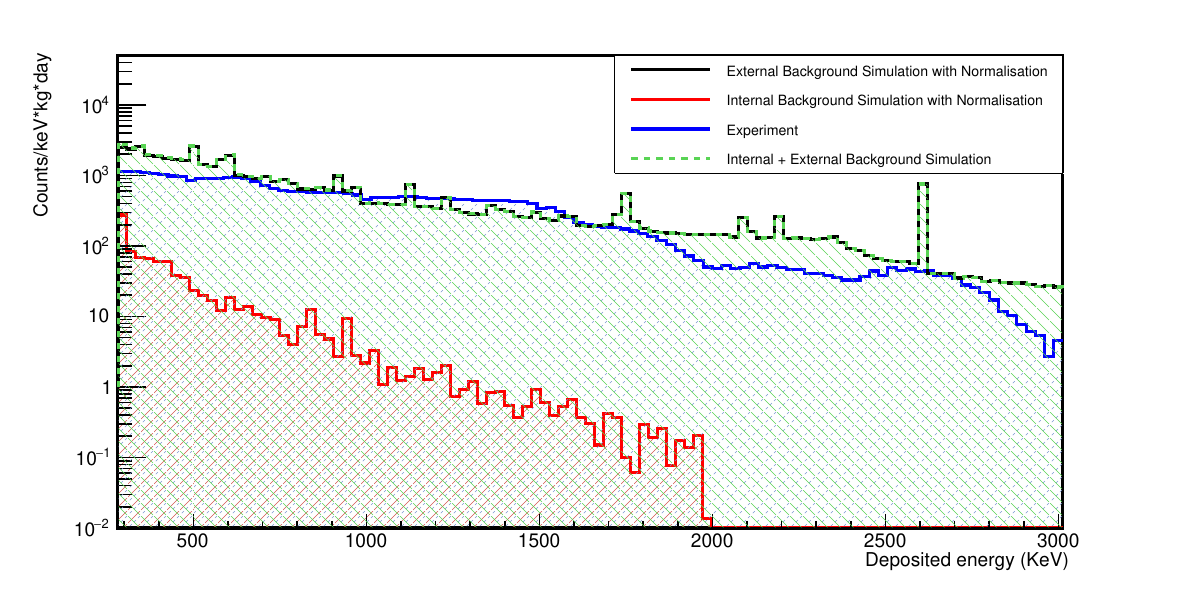}
   \caption{Comparison plot of simulated results with experiment. The y-axis is in DRU (differential-rate-unit). The x-axis is the energy deposited within the central CsI(Tl) crystal for a $3\times3$ geometry \cite{Verma}.}
  \label{fig:final comparison}
\end{figure}

The measured background spectrum from the $3\times3$ setup was also compared with a simulated spectrum. The background model was constructed using the SNOLAB $\gamma$-ray spectra corresponding to three primary sources: uranium, thorium, and neutrons, which together represent the dominant external background contributions. In the simulation, $\gamma$ rays were generated isotropically from the surface of a sphere surrounding the $3\times3$ detector geometry to mimic environmental radiation. Figure~\ref{fig:final comparison} shows the comparison between the experimental data and the simulated spectrum, demonstrating good agreement~\cite{Verma}. Since the \textsc{GEANT4} simulation does not account for real acquisition time, a normalization factor was applied to scale the simulated spectrum to the experimental data. The resulting agreement validates the background model used in this study.

\subsection{$5\times5$ CsI(Tl) Setup}

Following the successful operation of the $3\times3$ setup, a larger detector assembly consisting of 25 CsI(Tl) crystals arranged in a $5\times5$ geometry was constructed (see Fig.~\ref{fig:Detector}), increasing the fiducial mass from approximately 3.5~kg to 31.5~kg. The same procedures were followed with respect to shielding, electronics, data acquisition, and data analysis. In addition, a thermal management system was implemented to cool the photomultiplier tubes (PMTs), mitigating performance degradation and potential damage due to heating.

\begin{figure}[h!]
  \centering
    \includegraphics[scale=0.35]{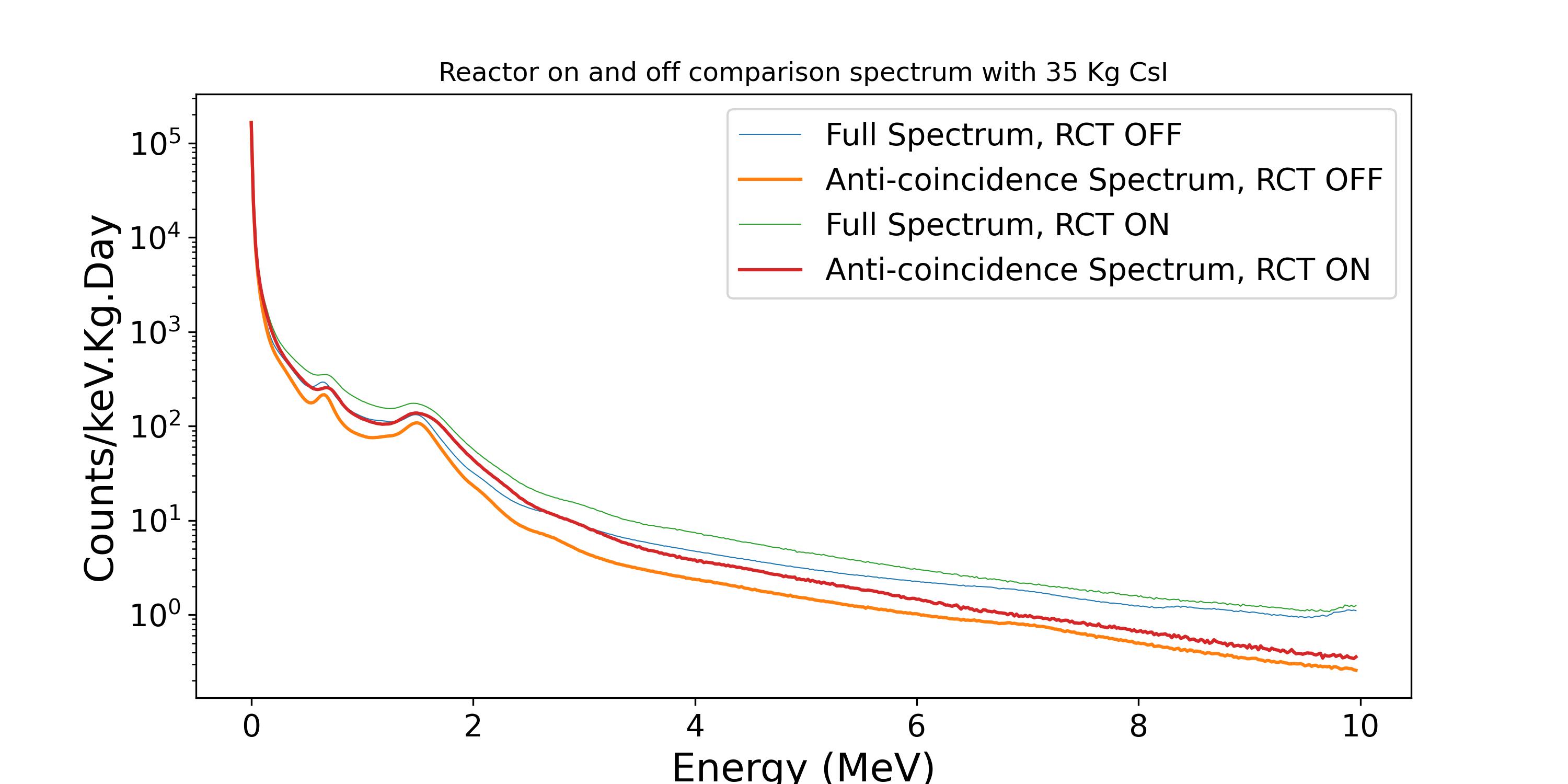}
   \caption{Reactor ON and OFF energy spectra obtained using the $5\times5$ CsI(Tl) setup. Sub-100 DRU background levels and clear reactor ON/OFF separation are observed above the 2~MeV region.}
  \label{comparison_new1}
\end{figure}

Figure~\ref{comparison_new1} shows the reactor ON and reactor OFF energy spectra obtained from the combined $3\times3$ central CsI(Tl) crystals, corresponding to a fiducial mass of 31.5~kg within the $5\times5$ detector assembly. The thin lines represent the full energy spectrum, while the thick lines correspond to the single-scatter (anti-coincidence) spectrum. The application of the single-scatter (active veto) technique reduces the background by a factor of approximately 2--3, achieving background levels of $\sim 0.1$--10~DRU in the energy range of 2--10~MeV~\cite{Verma}.

\section{Conclusion And Experimental Reach}

\begin{figure}[h!]
  \centering
    \includegraphics[scale=0.32]{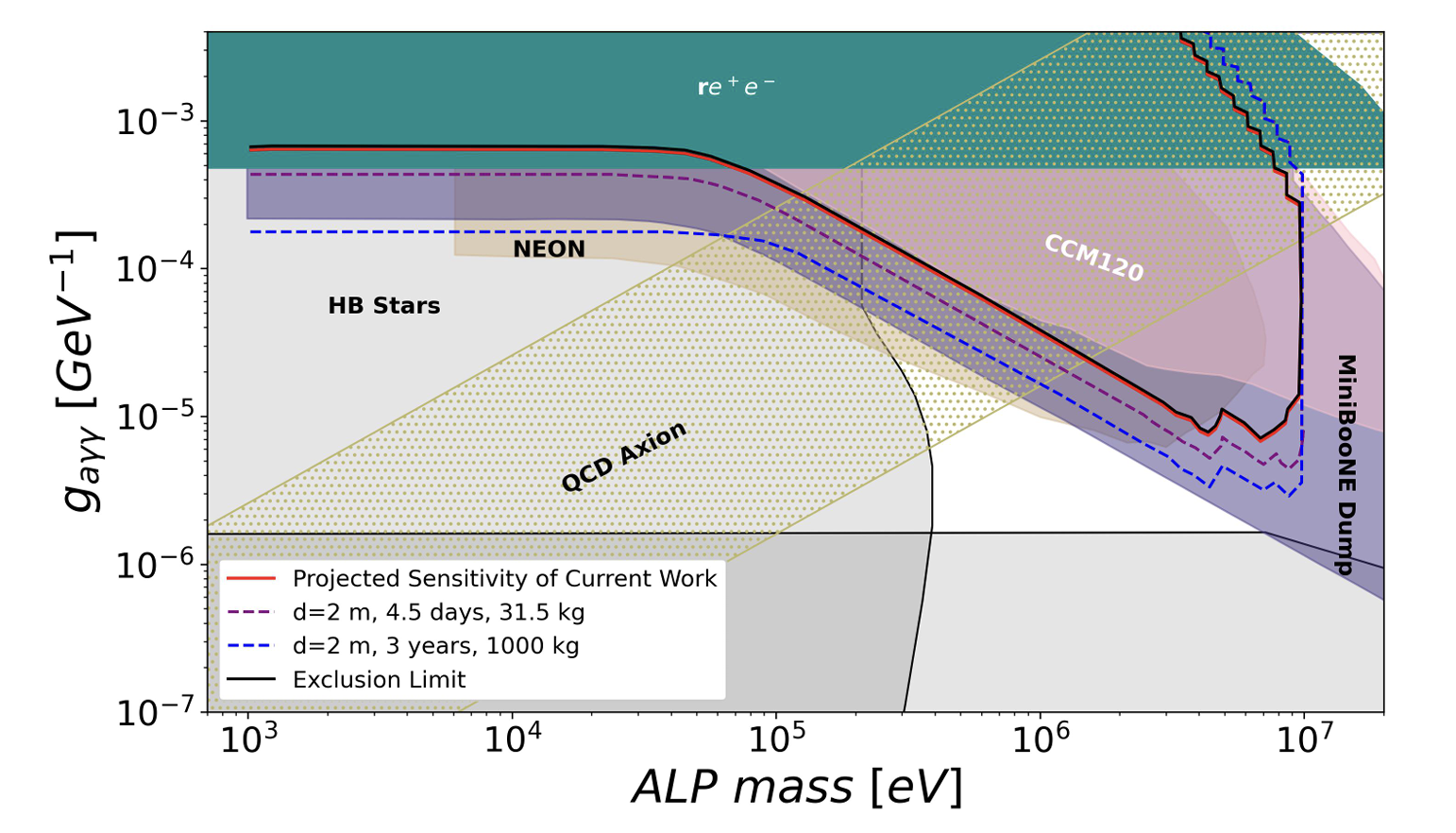}
   \caption{Exclusion plot drawn with the current experimental results. Red lines shown the contour region for the sensitivity of the experiment and the blue line shows the contour region for the excludes region after the experimental data being collected for around 2 days. }
  \label{Exclusion_plot}
\end{figure}

We calculate the projected upper limits on ALP couplings using a single-energy-bin analysis with the test statistic $\kappa = N_s^2 / N_b$, where $N_s$ and $N_b$ denote the integrated signal and background events, respectively. The upper limit on $N_s$ is obtained by setting $\kappa = 4.61$, corresponding to a $90\%$ confidence level (CL), and by assuming a background equal to the single-scatter event rate measured in reactor-off mode (see Fig.~\ref{Exclusion_plot}).
The figure shows the projected sensitivity of the experiment in the $(m_a, g_{a\gamma\gamma})$ parameter space, compared with the current exclusion limits. The results are obtained using the $5\times5$ detector configuration, corresponding to a fiducial mass of approximately $31.5~\mathrm{kg}$, with a dataset collected over 2 days at a distance of $4~\mathrm{m}$ from the reactor core.

The sensitivity curves are shown for different exposure times. For low ALP masses ($m_a \lesssim 4 \times 10^{4}~\mathrm{eV}$), the limits exhibit a nearly flat behavior, characteristic of inverse Primakoff scattering, which is dominated by forward scattering with negligible momentum transfer in this regime. At higher masses, the sensitivity becomes governed by the $a \rightarrow \gamma\gamma$ decay probability, which depends on the decay width (see Eq.~\ref{eq:photon_decay}) and the source-to-detector distance, giving rise to the characteristic peak structure in the exclusion curve. The decay width scales as $\Gamma \propto g_{a\gamma\gamma}^2 m_a^3$, and for models where $g_{a\gamma\gamma} \propto m_a$, this leads to $\Gamma \propto m_a^5$. Consequently, the decay lifetime decreases rapidly with increasing mass. As a result, light ALPs tend to be long-lived and may decay outside the detector, leading to reduced sensitivity, while heavier ALPs are more efficiently probed through decay channels. Finally, the sensitivity terminates for $m_a \gtrsim 10~\mathrm{MeV}$ due to the sharp decline in the reactor $\gamma$-ray flux above $E_\gamma \sim 10~\mathrm{MeV}$ (refer to Fig.\ref{fig:Flux_Reactor_Photon}). The overall behavior of the exclusion limits is illustrated in Fig.~\ref{Exclusion_plot}.

Also shown are existing constraints from astrophysical observations and laboratory experiments. Astrophysical limits arise from supernova SN1987A and horizontal branch (HB) stars~\cite{BROCKWAY1996439, PhysRevD.33.897, PhysRevD.36.2211, PhysRevLett.113.191302, CARENZA2020135709, Lucente_2020}. Laboratory bounds are provided by beam-dump and collider experiments at higher masses, while low-mass limits are derived from helioscope experiments such as CAST and SUMICO~\cite{AristizabalSierra:2020rom}.

The projected sensitivity demonstrates that the current setup can probe previously unexplored regions of parameter space. In particular, operating the detector at a closer distance to the reactor core ($\sim2~\mathrm{m}$) significantly enhances the sensitivity. Furthermore, scaling the detector mass to $\mathcal{O}(1000~\mathrm{kg})$ and extending the data-taking period to approximately three years would allow the experiment to approach the so-called cosmological triangle region~\cite{Verma}.

\section{Acknowledgements}
The work presented in this thesis has been supported by LANL - TRIAD  and DOE Grant Nos DE-SC0020097, DE-SC0018981, DESC0017859,
and DE-SC0021051. We also acknowledge Mitchell Institute and the Nuclear Science
and Engineering Center for providing institutional support with lab space and other required facilities
to carry out experimentation.

\bibliographystyle{elsarticle-num}
\bibliography{axion}


\section{Appendices}

\subsection{Background sources \label{sec:Background_Appendices}}

\subsubsection{Environmental Radioactivity}

\begin{center}
\begin{tabular}{||c c c c||} 
 \hline
 Element & Half-life & Decay product & Maximum energy (MeV)  \\ [0.5ex] 
 \hline\hline
 $^{238}$U  & $10^9$ yr & $\alpha$ & 7.8  \\ 
  &  & $\beta^-$ & 5.48  \\
  &  & $\gamma$ & 1.76 (dominant)  \\[1ex] 
 \hline
 $^{232}$Th & $10^{10}$ yr & $\alpha$ & 8.95 \\
  &  & $\beta^-$ & 5 \\
  &  & $\gamma$ & 2.6 (dominant)  \\[1ex]
 \hline
 $^{40}$K & $10^{9}$ yr & $\beta^-$ & 1.31  \\
  &  & $\gamma$ & 1.46  \\[1ex] 
 \hline
\end{tabular}
\end{center}

Environmental $\gamma$ radiation that can contribute to the background in CsI(Tl) detectors originates from three main sources: primordial, cosmogenic, and anthropogenic radioactivity~\cite{heusser}. Primordial isotopes have half-lives of $\gtrsim 10^{9}$ years and are naturally present in the Earth's crust. Among these, $^{238}$U, $^{232}$Th, and $^{40}$K are the dominant contributors. 

The decay chains of $^{238}$U and $^{232}$Th produce numerous $\gamma$ rays with different energies and intensities before reaching stable daughter nuclei. In addition, the decay of $^{40}$K produces a characteristic 1460~keV $\gamma$ ray following electron capture. These $\gamma$ rays from the uranium and thorium decay chains, together with the 1460~keV line from $^{40}$K, constitute the primary sources of natural external background radiation relevant to this experiment. An important feature of natural radioactivity is that the maximum $\gamma$-ray energy is limited to approximately 2.6~MeV, originating from the $^{208}$Tl decay in the $^{232}$Th chain.

\subsubsection{Cosmic Rays}

Cosmic-ray muons constitute another potential background source. At sea level, the muon flux is approximately $1~\mathrm{muon/cm^2/min}$. Muons are minimum ionizing particles and deposit energy in matter at a rate of approximately $(dE/dx) \sim 2~\mathrm{MeV\,g^{-1}\,cm^{2}}$. Given the density of CsI(Tl) (4.51~g/cm$^3$), vertically incident muons traversing the detector array (approximately 30~cm of CsI(Tl)) would deposit roughly $\sim270$~MeV of energy. This energy deposition is far above the region of interest for the present study and typically results in saturated signals in the digitizer. 

Even for horizontally incident muons passing through approximately 5~cm of CsI(Tl), the deposited energy would be about $\sim45$~MeV, which also lies well above the relevant energy range. Consequently, cosmic-ray muons are not expected to represent a significant background source for this experiment.

\subsubsection{Internal Radioactivity of CsI(Tl)}

\begin{figure}[h!]
  \centering
    \includegraphics[width=\linewidth]{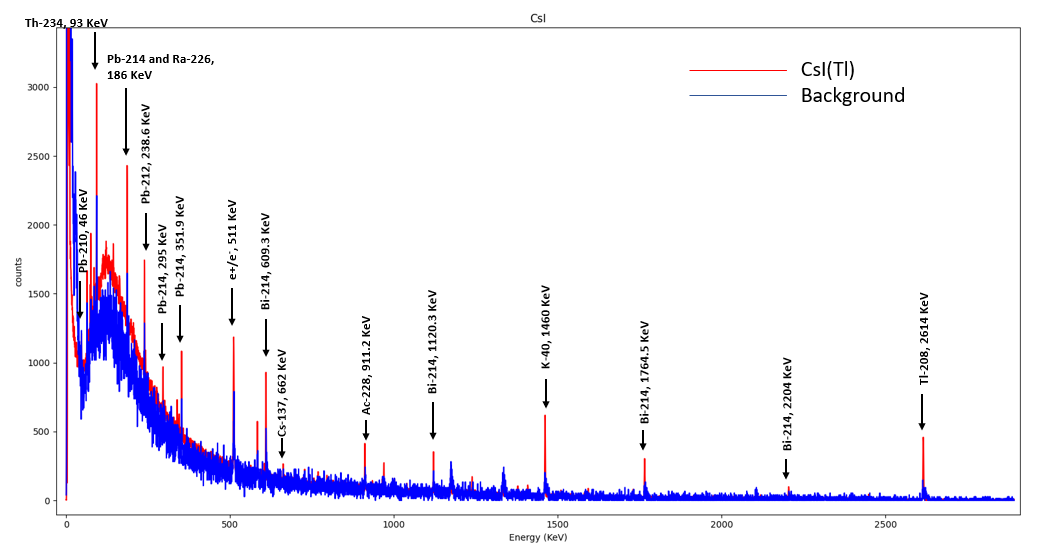}
   \caption{HPGe spectrum of the CsI(Tl) sample together with the background spectrum, showing the major identified $\gamma$-ray peaks~\cite{Verma}.}
  \label{fig:4}
\end{figure}

The internal radioactivity of the CsI(Tl) crystals has been characterized using a high-purity germanium (HPGe) detector, allowing the development of a reliable internal background model. Figure~\ref{fig:4} shows the HPGe spectrum measured for a CsI(Tl) sample together with the background spectrum, where the dominant $\gamma$-ray peaks are identified.

The major internal background contributions arise from $^{214}$Bi, $^{134}$Cs, $^{137}$Cs, and $^{40}$K. The 1460~keV $\gamma$ rays associated with $^{40}$K originate primarily from electron-capture decay~\cite{kim}. A fraction of this background may also arise from the surrounding lead shielding and concrete structures in the laboratory environment.

\end{document}